\documentclass[twocolumn,aps,pra,linenumber,showpacs,floatfix,superscriptaddress]{revtex4-1}
\usepackage{amsfonts}
\usepackage{amsmath}
\usepackage{amssymb}
\usepackage{dcolumn}
\usepackage{graphicx}%
\usepackage{bm}
\usepackage{color}
\newcommand{\braket}[3]{\left\langle #1 \middle| #2 \middle| #3 \right\rangle}

\begin{document}

\newcommand{\NZIAS}{
Centre for Theoretical Chemistry and Physics,
The New Zealand Institute for Advanced Study,
Massey University Auckland, Private Bag 102904, 0745 Auckland, New Zealand}

\newcommand{\UNSW}{
School of Physics, University of New South Wales, Sydney 2052, Australia}

\newcommand{\MBU}{
Department of Chemistry, Faculty of Natural Sciences, Matej Bel University, 
Tajovsk\'{e}ho 40, SK-974 00 Bansk\'{a} Bystrica, Slovakia}

\newcommand{\PUM}{Fachbereich Chemie, Philipps-Universit\"{a}t Marburg, 
Hans-Meerwein-Str., D-35032 Marburg, Germany}

\title[Anapole Moment]{The $P$--odd interaction constant $W_A$ from relativistic \textit{ab initio} calculations of diatomic molecules}


\author{A. Borschevsky}
\affiliation{\NZIAS}

\author{M. Ilia\v{s}}
\affiliation{\MBU}

\author{V. A. Dzuba}
\affiliation{\UNSW}

\author{K. Beloy}
\affiliation{\NZIAS}

\author{V. V. Flambaum}
\affiliation{\UNSW}

\author{P. Schwerdtfeger}
\affiliation{\NZIAS}
\affiliation{\PUM}

\date{\today}


\begin{abstract}
We present \textit{ab initio} calculations of the $W_A$ parameter of
the $P$-odd spin-rotational Hamiltonian for a variety of diatomic
molecules,  
including the group--2 and --12 halides. The results were obtained by
 relativistic Dirac--Hartree--Fock and density functional theory approaches, and corrected for core polarization effects. Strong enhancement of $W_A$ is found for the group--12
diatomic halides, which should be helpful in future determination 
of the nuclear anapole moment.
\end{abstract}

\maketitle

\section{Introduction}

The anapole moment was predicted first by Zeldovich \cite{Zel58} in 1958 as a new parity violating (PV) and time reversal (T) 
conserving moment of an elementary particle. It  appears in the
second-order multipole expansion of the magnetic vector-potential
simultaneously  
with the $P$-- and $T$-- violating  magnetic quadrupole moment
\cite{SusFlaKri84}. The nuclear anapole moment was experimentally
discovered  
in the $^{133}$Cs atom in 1997 \cite{WooBenCho97}. This measurement was performed following a proposal by Flambaum and
Khriplovich \cite{FlaKri80}, who have shown that the nuclear anapole provides the dominant contribution to the 
nuclear-spin-dependent (NSD) parity violating effect in atoms and molecules. It can provide important information 
about hadronic weak coupling, which is currently not so easily obtained from first-principles nuclear structure 
calculations (e.g., see Ref.~\cite{HaxLiuRam02} and Review \cite{GinFla04}).
The term in the Hamiltonian operator arising from NSD parity violating electron-nucleus interaction is 
\begin{eqnarray}
H_A=\kappa_{NSD} \frac{G_F}{\sqrt{2}}\frac{\bm{\alpha}\cdot \mathbf{I}}{I}\rho(\mathbf{r}),
\label{Ha}
\end{eqnarray}
where  $\kappa_{NSD}$ is the  dimensionless strength constant, $G_{F}=2.22249\times10^{-14}$ a.u. is the Fermi
constant, $\bm{\alpha}$ is a vector comprised of the conventional Dirac matrices, $\mathbf{I}$ is the
nuclear spin, $\mathbf{r}$ is the displacement of the valence electron from
the nucleus, and $\rho(\mathbf{r})$ is the (normalized) nuclear density. 
There are three sources for this interaction:
the first contribution arises from the electroweak neutral coupling between electron vector and 
nucleon axial-vector currents ($\mathbf{V}_e\mathbf{A}_N$) \cite{NovSusFla77}. The second contribution comes from the nuclear-spin-independent weak interaction combined with the hyperfine interaction \cite{FlaKri85}. 
Finally, the nuclear anapole moment contribution scales with the number of nucleons, $A$, with $\kappa_{A}\sim A^{2/3}$, 
and becomes the dominant contribution in spin-dependent atomic parity violation effects for sufficiently large nuclear charge $Z$ \cite{FlaKri80,FlaKriSus84}.
It requires nuclear spin $I\neq 0$ and 
in a simple  valence  model  has the following value   \cite{FlaKriSus84}
\begin{eqnarray}
\kappa_A= 1.15\times 10^{-3} \left(\frac{\mathcal{K}}{I+1} \right) A^{2/3} \mu_i g_i,
\label{eqanapole}
\end{eqnarray}
Here, $\mathcal{K}=(-1)^{I+\frac{1}{2}-l}(I+1/2)$, $l$ is the orbital
angular momentum of the external unpaired nucleon 
$i=n,p$;
 $\mu_p= +2.8$,
$\mu_n= -1.9$. Theoretical estimates give the strength constant for
nucleon-nucleus weak potential $g_p \approx +4.5$ for a proton and
$|g_n|\sim 1$ for a neutron \cite{FlaMur97}. The aim of anapole
measurements is to provide  accurate values for these constants. 
The nuclear anapole moment for $^{133}$Cs ($I$=7/2),
containing a single valence proton, was measured from the differences in the 
$6S_{F=4}$ to $7S_{F=3}$ and $6S_{F=3}$ to $7S_{F=4}$ hyperfine
transitions as $\kappa_A$=364(62)$\times$10$^{-3}$
\cite{WooBenCho97,FlaMur97}.  
However, the limit on $g_p$ ($g_p=-2 \pm 3$ \cite{GinFla04}),
obtained  from the Tl anapole measurements \cite{VetMeeMaj95}, seems to  
contradict the Cs anapole measurements
($g_p=6 \pm 1$, see \cite{FlaMur97}). 
This  
indicates that further refinements in the experimental measurements are required to obtain high precision
results for nuclear spin-dependent parity violation effects in atoms.

In Refs. \cite{Lab78,SusFla78,FlaKri85_2} it was shown that the
nuclear spin-dependent parity violation effects are enhanced by a
factor of $10^5$ in diatomic molecules with  $^2\Sigma_{1/2}$ and
$^2\Pi_{1/2}$ electronic states due to the mixing of close rotational
states of opposite parity ($\Omega$-doublet for
$^2\Pi_{1/2}$). DeMille and co-workers suggested therefore to measure
the anapole moment by using diatomic molecules in a  
Stark interference experiment to determine the mixing between
opposite-parity rotational/hyperfine levels \cite{DemCahMur08}. The
molecular route opens up the range of systems to be studied and should
provide data on anapole 
moments for many heavy nuclei. Another motivation comes from a possibility to test the standard  
model. The anapole
moment contribution is small in light nuclei with a valence neutron (Eq.~(\ref{eqanapole})). In  
this case the electroweak contribution
may be extracted from the measurements of NSD PV effects \cite{DemCahMur08}. This  
contribution has never been measured.
We therefore present Dirac-Hartree-Fock (corrected for electron 
correlation) and 4-component density functional theory 
calculations of the electronic 
factor $W_{A}$ for the diatomic group-2 and -12 fluorides and a number of 
other diatomic compounds.


\section{Computational Details.}
For $^2\Sigma_{1/2}$ and $^2\Pi_{1/2}$ electronic states, the interaction (\ref{Ha}) can be replaced by the effective operator, 
which appears in the spin-rotational Hamiltonian \cite{FlaKri85_2,DemCahMur08},
\begin{eqnarray}
H_A^\mathrm{eff}=\kappa_{NSD} W_A\frac{(\mathbf{n}\times\mathbf{S}^\prime)\cdot \mathbf{I}}{I},
\label{eq:Heff}
\end{eqnarray}
where $\mathbf{S}^\prime$ is the effective spin (discussed below) and
$\mathbf{n}$ is the unit vector directed along the molecular axis  
from the heavier to the lighter nucleus.

Calculations of the $P$-odd interaction constant $W_{A}$
were carried out within an open-shell single determinant
average-of-configuration Dirac-Hartree-Fock approach (DHF)
\cite{Thyssen_thesis} and within the relativistic density functional theory (DFT) \cite{SauHel02}, employing quaternion  
symmetry \cite{Saue:1997, Saue:1999}. We used the DIRAC10 computational package \cite{DIRAC10} to perform all the calculations. 
The electronic factor $W_A$ is found from evaluating the matrix
elements of the $\bm{\alpha}\rho(\mathbf{r})$ operator in the
molecular spinor basis \cite{Visscher1997181}. 
$^2\Sigma_{1/2}$ and $^2\Pi_{1/2}$ open-shell electronic states are
two-fold degenerate, corresponding to the two possible projections of
electronic angular momentum along $\mathbf{n}$,
i.e.~$|\Omega\rangle=|\pm\frac{1}{2}\rangle$.  
When operating within this degenerate space, the operator
$\frac{G_F}{\sqrt{2}}\bm{\alpha}\rho(\mathbf{r})$ is equivalent to
$W_A(\mathbf{n}\times\mathbf{S}^\prime$) (Eq.~(\ref{eq:Heff})).  
Time-reversal symmetry ensures that only the matrix elements that are
off-diagonal in $\Omega$ are non-vanishing. This symmetry rule is
encapsulated within the effective operator $H_A^\mathrm{eff}$ by the
angular factor $(\mathbf{n}\times\mathbf{S}^\prime)$. Here the
effective spin $\mathbf{S}^\prime$ generates rotations in the
degenerate subspace analogously to usual spin operator $\mathbf{S}$ in
a spin-1/2 system. In the non-relativistic limit,
$\mathbf{S}^\prime\rightarrow\mathbf{S}$ for $^2\Sigma_{1/2}$
states. A finite nucleus, modelled by the Gaussian 
charge distribution was employed \cite{VisDya97}. We note that a nuclear point charge approximation 
should be avoided in calculations of $W_{A}$ as the resulting singularity in the 
wave function gives unphysical results.

In the DFT calculations we used the Coulomb-attenuated B3LYP functional, 
(CAMB3LYP*), the parameters
of which were adjusted by Thierfelder \textit{et al.} \cite{ThiRauSch10} to reproduce the PV energy shifts obtained
using coupled cluster calculations. The newly adjusted parameters are $\alpha= 0.20$,
$\beta=0.12$, and $\mu=0.90$. In order to test the stability of the results with respect to the choice of the functional, the calculations were performed also using the PBE, LDA, and B3LYP functionals. For all the systems the $W_A$ parameters obtained using the different functionals were in remarkably good agreement, within up to $\sim10\%$ (the only exception being RaF, where the LDA results were $15\%$ higher than the CAMB3LYP* ones). We thus only present the CAMB3LYP* values, which are considered to give the best results for parity violating properties \cite{ThiRauSch10}.

For the lighter elements (N, O, F, Mg, and Cl), uncontracted aug-cc-pVTZ basis sets were
used \cite{KenDunHar92,WooDun93}. For the rest of the atoms, we employed
Faegri's dual family basis sets \cite{Fae01}. A good description of the electronic
wave function in the nuclear region is essential for obtaining reliable
results for parity violating properties \cite{LaeSch99}. Thus, we
augmented the basis sets with high exponent $s$ and $p$ functions, which
brings about an increase of around $10\%$ in the calculated values of $W_{A}$.
The basis sets were increased, both in the core and in the valence regions, to
convergence with respect to the calculated $W_{A}$ constants. The final basis sets can be found in the Appendix.

Where available, we used experimental geometries. For molecules where the bond
length $R_{e}$ is not known experimentally, we optimized the bond
distance instead, using relativistic coupled cluster
theory with single, double, and perturbative triple excitations
(CCSD(T)) \cite{Visscher:1996}. To reduce the computational effort,
the Dirac-Coulomb 
Hamiltonian was replaced by an infinite order two-component relativistic
Hamiltonian obtained after the Barysz--Sadlej--Snijders (BSS) transformation of
the Dirac Hamiltonian in a finite basis set \cite{IliJenKel05,Ilias:2007}. 
Our calculated $R_{e}$ are typically within 0.01 \AA \ of the experimental values, where available.
The experimental/calculated equilibrium distances
can be found in Tables \ref{tab:I} and \ref{tab:II}.

\begin{table*}
  \caption{Internuclear distances $R_{e}$ (\AA ) and the $P$-odd interaction constants $W_{A}$ (Hz) obtained on different levels of theory: DHF, DFT, DHF and DFT scaled for core polariration contribution, DFT$\cdot K_{CP}$ and DHF$\cdot K_{CP}$, DHF corrected for total correlation effect, DHF$\cdot K_{tot}$, and the final recommended values, taken as $W_A$(Final)$=(W_A$(DFT)$\cdot K_{CP}+W_A$(DHF)$\cdot K_{CP}$))/2. Comparison with earlier results is also shown.}
  \label{tab:I}
  \setlength{\tabcolsep}{0.3cm}
  \centering
  \resizebox{\textwidth}{!}{
  \begin{tabular}
    [c]{cccrrrrrrlrlll}\hline\hline
    & Nucleus & $R_{e}$ (\AA ) & \multicolumn{5}{c}{$W_{A}$ (Hz)} &\qquad&
    \multicolumn{3}{c}{Previous results}\\\cline{4-9}\cline{11-13}%

    \multicolumn{1}{l}{} & \multicolumn{1}{l}{} &
    \multicolumn{1}{l}{} &DFT&DHF&DFT$\cdot K_{CP}$&DHF$\cdot K_{CP}$&DHF$\cdot K_{tot}$&Final&& $W_{A}$ (Hz) & Method & Ref.\\\hline
    
    \multicolumn{1}{l}{SrF} & \multicolumn{1}{l}{Sr} & \multicolumn{1}{l}{2.075
      \cite{HubHerNIST}}&42 & 41 & 53 & 51 & 49 &52& & 65 &
    Semiempirical & \cite{DemCahMur08}\\

    \multicolumn{1}{l}{MgBr} & \multicolumn{1}{l}{Br} & \multicolumn{1}{l}{2.360
      \cite{HubHerNIST}} &18& 9 & 24&11&11& 18&&18 &
    Semiempirical & \cite{DemCahMur08}\\

    \multicolumn{1}{l}{ZnN} & \multicolumn{1}{l}{Zn} & \multicolumn{1}{l}{1.696
      \cite{HubHerNIST}} &56 & 63 &67 &76&70&72&& 99 &
    Semiempirical & \cite{DemCahMur08}\\

    \multicolumn{1}{l}{BaF} & \multicolumn{1}{l}{Ba} & \multicolumn{1}{l}{2.162
      \cite{HubHerNIST}}  & 121&123& 152&154& 142&153 && 164 &
    Semiempirical & \cite{DemCahMur08}\\
    \multicolumn{1}{l}{} & \multicolumn{1}{l}{} &
    \multicolumn{1}{l}{} &  & & & & & && 135 & DHF & \cite{NayDas09}\\
    \multicolumn{1}{l}{} & \multicolumn{1}{l}{}  &
    \multicolumn{1}{l}{} &  & & & & & && 160 & 4c-RASCI\footnote{Fully relativistic restricted active space configuration
      interaction method.} & \cite{NayDas09}\\
    \multicolumn{1}{l}{} & \multicolumn{1}{l}{}  &
    \multicolumn{1}{l}{} &  & & & & & && 111 & RECP+

SCF+EO\footnote{Relativistic effective core potential (RECP) combined with SCF 
      and an effective operator to account for core-valence correlations.} &
    \cite{KozTitMos97}\\
    \multicolumn{1}{l}{}  & \multicolumn{1}{l}{} &
    \multicolumn{1}{l}{} &  & & & & &  && 181 & RECP+RASSCF+EO\footnote{
      RECP combined with restricted active 
      space SCF approach and an effective operator to 
      account for core-valence correlations.} &
    \cite{KozTitMos97}\\
    \multicolumn{1}{l}{} & \multicolumn{1}{l}{} &
    \multicolumn{1}{l}{} & & & & &  &  && 210-240 & Semiempirical & \cite{KozLab95}\\
    \multicolumn{1}{l}{LaO} &
     \multicolumn{1}{l}{La} & \multicolumn{1}{l}{1.825
      \cite{HubHerNIST}} &161 & 164.3 & 197&201&186 &199 && 222 &
    Semiempirical & \cite{DemCahMur08}\\

    \multicolumn{1}{l}{YbF} & \multicolumn{1}{l}{Yb} & \multicolumn{1}{l}{2.016
      \cite{HubHerNIST}} &631 & 527& 657& 549& 590&602 && 729 &
    Semiempirical & \cite{DemCahMur08}\\
    \multicolumn{1}{l}{} & \multicolumn{1}{l}{}  &
    \multicolumn{1}{l}{} &  & &&&& && 484 & RECP+SCF & \cite{TitMosEzh96}\\
    \multicolumn{1}{l}{}  & \multicolumn{1}{l}{} &
    \multicolumn{1}{l}{} &  &  &&&&&& 486 & RECP+RASSCF & \cite{TitMosEzh96}\\

    \multicolumn{1}{l}{HgF} & \multicolumn{1}{l}{Hg} & \multicolumn{1}{l}{2.025\footnote{CCSD(T), present calculations}}  &3207& 3557&3502&3884 &3023& 3693 && 2700 & Semiempirical &
    \cite{KozLab95}\\

    \multicolumn{1}{l}{PbF} & \multicolumn{1}{l}{Pb} & \multicolumn{1}{l}{2.078
      \cite{HubHerNIST}} &$-1382$ & $-1349$ &$-1517$&$-1481$&$-1200$&$-1499$ && $-720$ &
    Semiempirical & \cite{DmiKhaKoz92}\\
    \multicolumn{1}{l}{} & \multicolumn{1}{l}{}  &
    \multicolumn{1}{l}{} &  & & & & & && $-950\pm$300 & Semiempirical &
    \cite{KozFomDmi87}\\

    \multicolumn{1}{l}{}  & \multicolumn{1}{l}{} &
    \multicolumn{1}{l}{} &  & & & & & && $-990$ & RECP+SODCI\footnote{
      RECP combined with spin-orbit direct configuration
interaction} &
    \cite{BakPetTit10}\\

    \multicolumn{1}{l}{RaF} & \multicolumn{1}{l}{Ra} & \multicolumn{1}{l}{2.255\footnotemark[4]} &1681 & 1465 & 2054& 1790& 1773& 1922 & & 1300 &
    ZORA+SCF\footnote{Quasirelativistic two-component zero-order regular approximation
      combined with the SCF approach.} & \cite{IsaHoeBer10}\\\hline
  \end{tabular}
}
\end{table*}

Previous investigations for the BaF molecule have shown that the electron
correlation contribution to the $W_{A}$ constant is non-negligible,
and raises its value by $\sim$20-50\% \cite{NayDas09,KozTitMos97}.
In this work we use two separate schemes to account for the correlation effects: the density functional calculations, and correcting the DHF values for the correlation contributions using atomic caculations, in a manner outlined below.

The main 
contribution to the matrix elements of the NSD interaction for the 
valence molecular electrons comes from short distances around the heavy nucleus. 
Thus, these matrix elements are strongly affected by correlations 
between the core and the valence electrons. 

The total molecular potential at short distances from the heavy nucleus
is spherically symmetric to very high precision; the core of the heavy
atom is practically unaffected by the presence of the second atom. The molecular orbitals of the valence 
electron can thus be expanded in this region, using spherical harmonics centered at
the heavy nucleus,
\begin{equation}
  |\psi_v \rangle= a |s_{1/2} \rangle +b |p_{1/2} \rangle + c|p_{3/2} \rangle + d|d_{3/2} \rangle \dots
\label{eq:psi_v}
\end{equation}
Only $s_{1/2}$ and $p_{1/2}$ terms of this expansion give significant
contribution to the matrix elements of the weak interaction.
These functions can be considered as states of an atomic
 valence electron, and are calculated
using standard atomic techniques in two different approximations: one
that includes electron correlation and another that does not. 
The correlation factor $K_{tot}$ is then defined as $K_{tot} = K_W \cdot K_{E1} \cdot K_{E_n}$. 
Here, $K_W$
is found as the ratio of the matrix elements
\begin{equation}
  K_{W} = \frac{\langle ns^{\prime}_{1/2}| \hat H_{\rm A}^{\prime}
    |np^{\prime}_{1/2} \rangle}{\langle ns_{1/2}|
    \hat H_{\rm A} |np_{1/2} \rangle}
\label{eq:K0}
\end{equation}
where the matrix element in the numerator includes electron correlation, while
the matrix element in the denominator does not. 
The magnitude of $K_W$ is larger than 1,
as the many body corrections due to correlation with the core
electrons increase the density of the valence electron on
the nucleus \cite{BanHay71}, and thus, increase the $W_{A}$ constant.

Correlations also
modify the expansion coefficients $a$, $b$, $c$... in
Eq.~(\ref{eq:psi_v}). An estimate of this effect shows that it reduces
the overall correlation contribution, and provides us with $K_{E1}$ and $K_{E_n}$, corresponding to the effect of correlations on the $E1$ amplitude and on the orbital energies, respectively. 

 Multiplying the result of molecular DHF calculations by $K_{tot}$ allows us to include the effect
of the most important electron correlation contributions (Table I, DHF$\cdot K_{tot}$). The Appendix contains the details of the calculation and the values of $K_{tot}$ for all the atoms studied here. 

Core polarization effects are not accounted for in the Kramers 
restricted DFT calculations. These effects are included in $K_{tot}$; however, we also examine their influence separately from the rest of the correlation contributions. Thus, we define an additional scaling factor, $K_{CP}$, which only takes the core polarization into account (see Appendix for the details of calculation and the values of $K_{CP}$). Table I contains the DHF and the DFT values, corrected for core polarization contribution (DHF$\cdot K_{CP}$ and DFT$\cdot K_{CP}$).

As the recommended value for the $W_A$ parameter we take an average of $W_A$(DHF)$\cdot K_{CP}$ and $W_A$(DFT)$\cdot K_{CP}$, since we believe that these two values represent our most reliable  
results.

\section{Results and Discussion}

 All the systems under study have a single valence electron; the ground state
of the PbF molecule is $^{2}\Pi_{1/2}$, the remaining molecules have a
$^{2}\Sigma_{1/2}$ ground state. Table I contains the results obtained for
molecules, for which earlier calculations were performed. For comparison, we present the uncorrelated DHF results, together with the different correlation schemes used: the DFT values, the DHF and the DFT values corrected for core polarization, the DHF values corrected for the overall correlation effects, $W_A$(DHF)$\cdot K_{tot}$, and the final recommended values. To the best of our knowledge, these are the first DFT calculations of $W_A$. The spread of the values obtained by different methods may be used to estimate the accuracy of our calculations, as the uncertainty of the result is strongly dominated by the correlation  
correction contribution. The difference between the DFT and DHF values gives us an estimate for  
the correlation contribution, which is included into DFT and absent in  DHF.  Another (and probably  
less accurate) method to estimate the correlations is to compare $K_{CP}$ and $K_{tot}$. From these  
comparisons we have come to conclusion that the accuracy of $W_A$ produced by the metal anapole moments is  
about 15\% for $\Sigma$ terms and 20-30\% for $\Pi_{1/2}$ terms (such as PbF). The accuracy of $W_A$ produced by   
Br (or other halogen) anapole moment is about 30\%.

\begin{table}
  \caption{Internuclear distances, $R_{e}$ (\AA ), relativistic factors $R_W%
$, and the recommended values of $W_{A}$ constants (Hz), taken as $W_A$(Final)$=(W_A$(DFT)$\cdot K_{CP}+W_A$(DHF)$\cdot K_{tot}$))/2 for group-2 and -12 fluorides, and for mercury halides.}
  \label{tab:II}
  \setlength{\tabcolsep}{0.3cm}
  \centering
  \resizebox{\columnwidth}{!}{
  \begin{tabular}[c]{lllrr}
    \hline\hline
           & $Z$& $R_{e}$ (\AA )& $R_W$ & $W_{A}$(Final) (Hz)\\\hline
           \multicolumn{5}{c}{Group-2 fluorides}\\*[0.1cm]
    MgF    & 12 & 1.750 \cite{HubHerNIST}& 1.06 &5\\
    CaF    & 20 & 1.967 \cite{HubHerNIST}& 1.17 &11\\
    SrF    & 38 & 2.075 \cite{HubHerNIST}& 1.65 &52\\
    BaF    & 56 & 2.162 \cite{HubHerNIST}& 2.73  &153\\
    RaF    & 88 & 2.244\footnote{CCSD(T), present calculations}& 10.34 &1922\\
               \multicolumn{5}{c}{Group-12 fluorides}\\*[0.1cm]
    ZnF & 30  & 1.766 \cite{FloMcLZiu06} & 1.38 &64\\
    CdF & 48  & 1.991\footnotemark[1]& 2.14 & 264\\
    HgF & 80  & 2.025\footnotemark[1]& 7.03 &3693\\
               \multicolumn{5}{c}{Mercury halides}\\*[0.1cm]
    HgCl & 80  & 2.387\footnotemark[1]& 7.03 & 3647\\
    HgBr & 80  & 2.468\footnotemark[1]& 7.03 &3600\\
    HgI & 80  & 2.736\footnotemark[1]& 7.03  &3356\\\hline
  \end{tabular}
  }
\end{table}

Most of the previous investigations of $W_A$ rely on semiempirical methods, while for
BaF, YbF, RaF, and PbF \textit{ab initio} calculations were also performed
\cite{TitMosEzh96,KozTitMos97,NayDas09,IsaHoeBer10,BakPetTit10}. 
Our results are in good agreement with 
the recent semiempirical values for most of the systems; in case of HgF, PbF, and RaF our final $W_A$ constants are higher than the previous values by about 30\%. 

For BaF we can compare our results to other \textit{ab initio} calculations. Our DHF value is close to that of Nayak
and Das \cite{NayDas09}, obtained by a similar method, and to that of Kozlov \textit{et al.} \cite{KozTitMos97},
calculated by the combination of relativistic effective core potential (RECP)
and the SCF approach, and corrected for valence-core correlation by an effective
operator. The value of $W_{A}$ corrected for correlation is in very good agreement with both the relativistic restricted active space configuration
interaction (4c-RASCI) result of Ref.~\cite{NayDas09}, and the RECP restricted active
space SCF approach of \cite{KozTitMos97}.   

Two other systems with \textit{ab initio} results are YbF
\cite{TitMosEzh96}, calculated using a combination of RECP and SCF/RASCF
methods, and RaF \cite{IsaHoeBer10}, treated via the quasirelativistic
two-component zero-order regular approximation (ZORA) combined with an SCF
approach. As both these investigations do not treat electron correlation, we can compare them directly to our 
DHF results, which are in good agreement.

Our correlated result for PbF is rather higher than that obtained in the previous investigation \cite{BakPetTit10} performed using a combination of RECP and spin-orbit direct configuration interaction, which might be caused by the 4-component treatment of relativity in our case. This system is different from the other molecules discussed here, due to its ground state. For the $^2\Pi_{1/2}$ electronic state $W_A$ vanishes in the non-relativistic limit, since in this limit it does not contain the $s$-wave electronic orbital and can not provide the matrix element $<s_{1/2}|W_A|p_{1/2}>$. The effect appears due to the mixing of  $^2\Sigma_{1/2}$ and
$^2\Pi_{1/2}$ electronic states by the spin-orbit interaction. This gives an extra factor of $Z^2 \alpha^2$
in the $Z$-dependence of the matrix element of $W_A$ in a $^2\Pi_{1/2}$ electronic state. This does not lead to a significant suppression in heavy molecules, such as PbF; however, in light molecules the matrix element of $W_A$ in the
$^2\Pi_{1/2}$ electronic state is much smaller than that in  $^2\Sigma_{1/2}$  state (note that a similar factor $Z^2 \alpha^2$ also makes the interval between the opposite parity $\Lambda$-doublet states small; therefore, there is actually no suppression in the mixing of these states by $W_A$).

Table \ref{tab:II} contains the recommended $W_{A}$ constants for the group-2 and group-12
fluorides. The magnitude of $W_{A}$ in $^2\Sigma_{1/2}$ electronic state is expected to scale as $Z^{2}R_W$ \cite{FlaKri85}, where $R_W$ is the relativistic parameter,
\begin{align}
R_W  & =\frac{2\gamma+1}{3}\left(  \frac{a_{B}}{2Zr_{0}A^{1/3}}\right)
^{2-2\gamma}\frac{4}{\left[  \Gamma(2\gamma+1)\right]  ^{2}},\label{Rw}\\
\gamma & =[1-(Z\alpha)^{2}]^{1/2}.\nonumber
\end{align}

\begin{figure}
  \centering
  \includegraphics[width=0.5\textwidth]{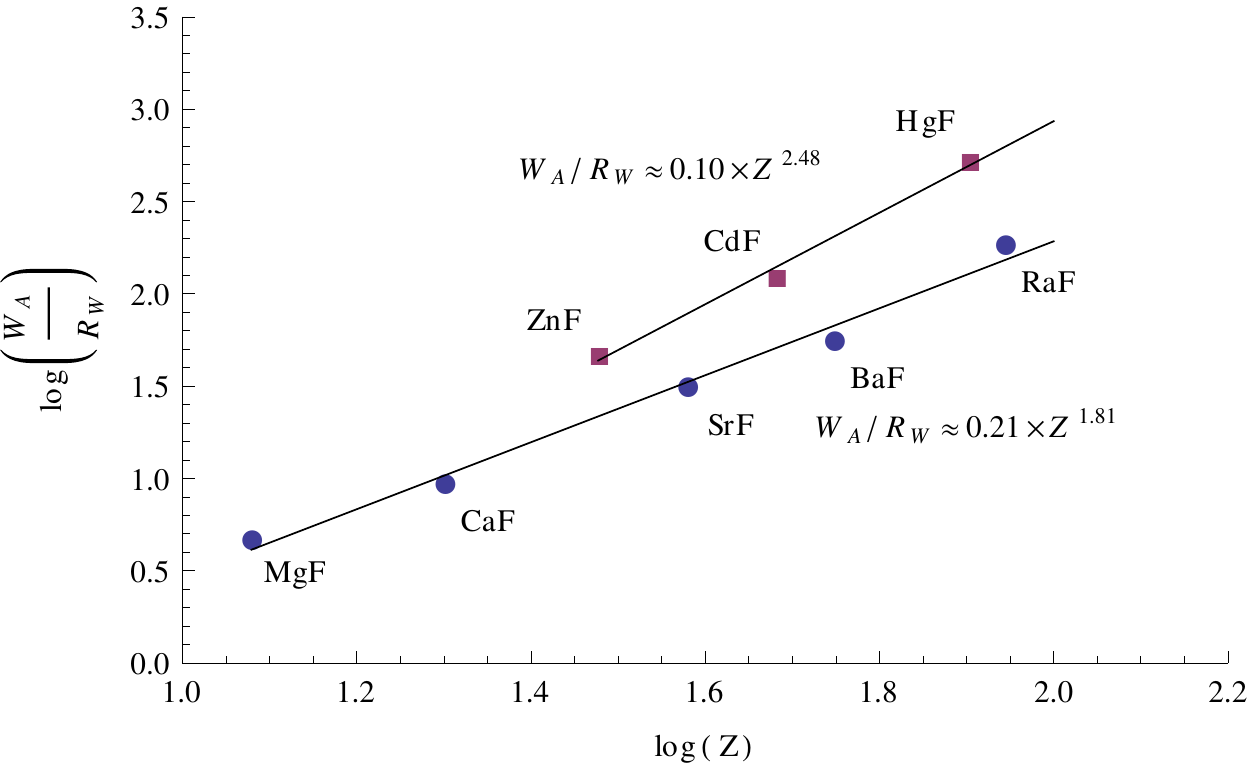}
  \caption{Scaling of $\log\left(
\frac{W_{A}}{R_{W}}\right)  $ with $\log(Z)$ for group-2 and -12 fluorides}
  \label{fig:I}
\end{figure}

In Eq.~(\ref{Rw}), $a_{B}$ is the Bohr radius, $r_{0}=1.2\times10^{-15}$
m, and $\alpha$ is the
fine-structure constant. The $R_{W}$ parameters are shown in Table
\ref{tab:II} for each of the metal atoms. In Fig. \ref{fig:I} we plot $\log\left(
\frac{W_{A}}{R_{W}}\right)  $ as a function of $\log(Z)$ for both groups of
dimers. For group-2 fluorides the scaling is, indeed, close to $Z^{2}$;
however, the interesting feature of the plot is group-12 fluorides, where the
$Z$-dependence is much more advantageous, of $Z^{2.5}$. This is due to the 
filling of the lower lying $d$-shell, which expands relativistically and thus 
increases the effective nuclear charge, leading
to an enhancement of relativistic and PV effects \cite{AutSieSet02}, and an increase of $W_{A}$. 
It should be noted that measurements for $W_A$ in MgF and CaF may be used to test  
the standard model, since the anapole moments of Mg and Ca
are small and the electroweak contribution to NSD PV electron-nucleus  
interaction is important.

Table \ref{tab:II} also shows the $W_{A}$ constants of mercury halides. The $W_{A}$ values are very close for all the halogens; the situation is similar for zinc and cadmium halides. However, molecules containing heavier ligands might have
an experimental advantage. Due to the higher reduced mass, the interval
between the opposite parity rotational levels becomes smaller, and thus larger
PV effects can be expected, and a smaller magnetic field would
be needed to reduce the interval between the levels.

\textbf{Acknowledgements}
This work was supported by the Marsden Fund (Royal Society of New Zealand), the Australian Research Council
and the Alexander von Humboldt Foundation (Bonn). MI is grateful for the financial support from the Slovak Research and 
Development Agency (grant number APVV-0059-10).

\section{Appendix}

\subsection{Basis sets}

Table \ref{tab:III} contains the basis sets used in the calculations of the $W_{A}$ constants. 
For the lighter elements (N, O, F, Mg, and Cl), uncontracted augmented
correlation-consistent valence triple-$\zeta$ (AVTZ) Gaussian basis sets were
used \cite{KenDunHar92,WooDun93}. For the rest of the atoms, we employed
the Faegri's dual family basis sets \cite{Fae01}, augmented with higher
orbital momentum and diffuse functions. Good description of the electronic
wave function in the nuclear region is essential for obtaining reliable
results for parity violating properties \cite{LaeSch99}. Thus, we also
augmented the basis sets with high exponent $s$ and $p$ functions, which
brings about an increase of around $10\%$ in the calculated values of $W_{A}$.
The basis sets were increased, both in the core and in the valence regions, to
achieve convergence with respect to the obtained $W_{A}$ constants.
The same basis sets were employed in the relativistic CCSD(T) geometry optimizations (where experimental geometry was unavailable). However, here we leave out the high exponent
$s$ and $p$ functions, as these contribute little to molecular geometries.

\subsection{Calculation of the correlation factors $K_{CP}$ and $K_{tot}$}



\begin{table}
  \caption{Basis sets employed in the calculation of the $W_{A}$ constants. All
elements with $Z>17$ are described by the Faegri basis sets \cite{Fae01}
augmented by high exponent, diffuse, and high angular momentum functions.}
  \label{tab:III}
  \centering
\begin{tabular}
[c]{lrr}\toprule
Atom\ \ \  &\ \ \ \  $Z$ &\hspace{2.5cm} Basis Set\\\colrule
N & 7 & aug-cc-PVTZ\\
O & 8 & aug-cc-PVTZ\\
F & 9 & aug-cc-PVTZ\\
Mg & 12 & aug-cc-PVTZ\footnote{augmented by 2 high exponent $s$ and 4 high exponent $p$ functions.}\\
Cl & 17 & aug-cc-PVTZ\\
Ca & 20 & 20\textit{s}18\textit{p}9\textit{d}6\textit{f}1\textit{g}\\
Zn & 30 & 21\textit{s}19\textit{p}10\textit{d}7\textit{f}2\textit{g}\\
Br & 35 & 21\textit{s}20\textit{p}10\textit{d}10\textit{f}1\textit{g}\\
Sr & 38 & 21\textit{s}20\textit{p}12\textit{d}9\textit{f}2\textit{g}\\
Zr & 40 & 21\textit{s}20\textit{p}12\textit{d}9\textit{f}2\textit{g}\\
Cd & 48 & 22\textit{s}20\textit{p}12\textit{d}9\textit{f}2\textit{g}\\
I & 53 & 22\textit{s}21\textit{p}12\textit{d}11\textit{f}2\textit{g}\\
Ba & 56 & 24\textit{s}22\textit{p}15\textit{d}10\textit{f}2\textit{g}\\
La & 57 & 24\textit{s}22\textit{p}14\textit{d}10\textit{f}2\textit{g}\\
Yb & 70 & 26\textit{s}21\textit{p}14\textit{d}10\textit{f}2\textit{g}\\
Hg & 80 & 25\textit{s}21\textit{p}15\textit{d}10\textit{f}2g\\
Pb & 82 & 25\textit{s}22\textit{p}16\textit{d}10\textit{f}2\textit{g}\\
Ra & 88 & 26\textit{s}23\textit{p}16\textit{d}11\textit{f}2\textit{g}%
\\\toprule
\end{tabular}
\end{table}
The scaling factors $K_{CP}$ and $K_{tot}$ are used for estimating the effect of the core polarization and of the overall correlation effects on $W_A$ for the molecules. These factors are found using two sets of atomic calculations, one that neglects electron correlation and on that includes it. The calculations, and the derivation of the $K_{CP}$ and $K_{tot}$ factors are described below.

\subsubsection{Atomic calculations without electron correlation}

We use the relativistic Dirac--Hartree--Fock method (DHF) to perform atomic
calculations \cite {DzuFlaSil87}. In atomic units ($|e|=1, \hbar=1, m_e=1$), the single electron DHF Hamiltonian is given by
\begin{equation}
  \hat H_0 = c \bm{\alpha} \cdot \mathbf{p} + (\bm{\beta} -1)c^2 -
  \frac{Z}{r} + V_e(r), 
\label{eq:h0}
\end{equation}
where $\bm{\alpha}$ and $\bm{\beta}$ are the Dirac matrices and $V_e(r)$ is the self-consistent DHF potential due
to atomic electrons. In order to take into
account the specifics of diatomic molecules, we use a slightly modified $V_e(r)$ potential 
compared to standard atomic techniques: 
\begin{equation}
  V_e(r) = V_{\rm DHF}^{N-N_v} + \frac{N_v-1}{\sqrt{r^2+R_e^2}}.
\label{eq:vtot}
\end{equation}
Here $V_{\rm DHF}^{N-N_v}$ is the self-consistent DHF potential of
the closed-shell core of the heavy atom, $N$ is total number of
electrons in this atom ($N=Z$ for a neutral system), $N_v$ is the
number of valence electrons, $r$ is the distance to the heavy nucleus,
and $R_e$ is the distance between the nuclei in the molecule. The second term
in (\ref{eq:vtot}) represents the spherically symmetric contribution
from the valence electrons that are assumed to be moved to the second atom. Its
form is chosen to have the correct $-1/r$ asymptote 
for the total potential at large distances in the case of neutral
molecule. 

The self-consistent DHF procedure is first done for the ion, from
which valence electrons are removed. Then the core potential $V_{\rm
  DHF}^{N-N_v}$ is frozen and valence $s_{1/2}$ and $p_{1/2}$ states
are calculated by solving the DHF equations for the valence electron
\begin{equation}
(\hat H_0 - \epsilon_v) \psi_v=0,
\label{eq:DHF}
\end{equation}
where $\hat H_0$ is given by (\ref{eq:h0}) and (\ref{eq:vtot}).

For example, in the case of the BaF molecule, $Z=N=56$, $N_v=2$, and $R_e=2.162$
\AA~\cite{HubHerNIST}. The self consistent DHF procedure is performed for the
Ba$^{2+}$ ion. The $6s_{1/2}$ and $6p_{1/2}$ valence states are calculated in
the potential 
\begin{equation}
  V(r) = -\frac{56}{r} + V_{\rm DHF}({\rm Ba}^{2+}) +
  \frac{1}{\sqrt{r^2+R_e^2}}. 
\label{eq:BaV}
\end{equation}
These $6s_{1/2}$ and $6p_{1/2}$ states are used in the denominator of (\ref{eq:K0}). 

\subsubsection{Inclusion of electron correlation}

We include two important classes of electron correlation corrections: the core
polarization correction and the Brueckner-type
correlations. These types of correlations dominate in
$s_{1/2}$ and $p_{1/2}$ atomic states and their inclusion leads to an accuracy of a few percent
for the matrix elements~\cite{DzuFlaSil87}.

The core polarization can be understood as the change of the
self-consistent DHF potential due to the effect of the extra term (the
weak interaction operator $\hat H_{\rm A}$) in
the Hamiltonian. The inclusion of the core polarization in
a self-consistent way is equivalent to the well-known random-phase
approximation (RPA) (see, e.g. \cite{DzuFlaSil87}). The change of the DHF
potential is found by solving the RPA-type equations self-consistently
for all states in the atomic core. The RPA equations have a form of the
DHF equations with the right-hand side:
\begin{equation}
(\hat H_0 - \epsilon_c) \delta \psi_c = (-\hat H_{\rm A} + \delta V_{\rm A})\psi_c
,
\label{eq:RPA}
\end{equation}
where $\hat H_0$ is the DHF Hamiltonian (\ref{eq:h0}), index $c$
enumerates states in the core, $\delta \psi_c$ is the correction to the
core state $c$ due to weak interaction $\hat H_{\rm A}$, and $\delta V_{\rm A}$
is the correction to the self-consistent core potential due to the
change of all core functions. Once $\delta V_{\rm A}$ is found, the core polarization can be
included into a matrix element for valence states $v$ and $w$ via the
redefinition of the weak interaction Hamiltonian,
\begin{equation}
\langle v |\hat H_{\rm A}| w \rangle \rightarrow \langle v |\hat H_{\rm
  A} + \delta V_{\rm A}| w \rangle. 
\label{eq:meRPA}
\end{equation}

We then obtain the scaling parameter for core polarization, $K_{CP}$, from

\begin{equation}
  K_{CP} = \frac{\langle \psi^{\rm DHF}_{ns_{1/2}}  | \hat H_{\rm A} + \delta
    V_{\rm A}
    |\psi^{\rm DHF}_{n^{\prime}p_{1/2}} \rangle}{\langle \psi^{\rm DHF}_{ns_{1/2}}|
    \hat H_A |\psi^{\rm DHF}_{n^{\prime}p_{1/2}} \rangle} .
\label{eq:K_CP}
\end{equation}

The values of $K_{CP}$ for all the atoms under study are presented in Table \ref{tab:IV}.

In contrast to the core polarization correction, which can be reduced
to the redefinition of the interaction Hamiltonian, the Brueckner-type
correlations can be reduced to the
redefinition of the single-electron orbitals, by replacing the DHF
orbitals by the Brueckner orbitals (BO). Brueckner
correlations describe the interaction between the valence and the
core electrons. These correlations can be
included with the use of the so-called correlation potential $\hat
\Sigma$, which is defined in such a way that the average value of $\hat
\Sigma$ over a valence state $v$ is the correlation correction to the
energy of this state:
\begin{equation}
 \delta \epsilon_v = \langle v| \hat \Sigma | v \rangle.
\label{eq:Sigma}
\end{equation}
The correlation potential $\hat \Sigma$ is a non-local
operator similar to the DHF exchange potential. It can be calculated
by means of the many-body perturbation theory (MBPT) in the residual
electron-electron Coulomb
interaction. The expansion starts from second order in this interaction and in most cases this is the leading term. We use the
B-splines in a box~\cite{JohSap86} and the second-order MBPT to
calculate $\hat \Sigma$. The Brueckner orbitals for the valence states
are found by solving the DHF-like equations with an extra operator
$\hat \Sigma$ included: 
\begin{equation}
(\hat H_0 +\hat \Sigma - \epsilon_v) \psi^{\mathrm {BO}}_v=0.
\label{eq:DHF2}
\end{equation}
Solving these equations gives new energies and new wave functions for
the valence states. For all atoms considered in present paper the
inclusion of the second-order correlation potential $\hat \Sigma$
leads to a few percent accuracy for the energies of the $s_{1/2}$ and
$p_{1/2}$ states.
Matrix element of the operator $\hat H_{\rm A}$ between valence
states $v$ and $w$, in which both type of correlations are included,
is given by
\begin{equation}
\langle \psi^{\mathrm{BO}}_v| \hat H_{\rm A} + \delta V_{\rm A} |\psi^{\mathrm{BO}}_w \rangle.
\label{eq:meBO}
\end{equation}
The $K_W$ factor (Eq. (\ref{eq:K0})) is then reduced to
\begin{equation}
  K_W =
\frac{\langle \psi^{\mathrm {BO}}_{ns_{1/2}}| \hat H_{\rm A} + \delta
    V_{\rm A}
    |\psi^{ \mathrm{BO}}_{n^{\prime}p_{1/2}} \rangle}{\langle \psi^{\rm DHF}_{ns_{1/2}}|
    \hat H_A |\psi^{\rm DHF}_{n^{\prime}p_{1/2}} \rangle} .
\label{eq:K}
\end{equation}

An additional effect of correlation, not taken into account in the above is the change in the expansion coefficients $a$, $b$, $c$... in Eq. (\ref{eq:psi_v}). In order to treat this effect we turn to the following expression,
\begin{equation}
  \frac{ \braket{ns_{1/2}}{H_A}{np_{1/2}} \braket{np_{1/2}}{E1}{ns_{1/2}} }{E_{ns_{1/2}} - E_{np_{1/2}}}
\label{eq:New}.
\end{equation}
\begin{table}
\caption{$K{CP}$, $K_W$, $K_{E1}$, $K_{E_n}$, and $K_{tot}$ correlation scaling vactors for the atoms under study.}
\label{tab:IV}
  \centering
\begin{tabular}{lllllc}
\hline\hline
Atom&$K_{CP}$\ \ \ \ \ &$K_W$\ \ \ \ \ &$K_{E1}$\ \ \ \ \ &$K_{E_n}$\ \ \ \ \ &$K_{tot}$\ \ \\
\hline
Mg&1.24&1.38&0.97&0.96&1.29\\
Ca&1.28&1.54&0.91&0.90&1.26\\
Zn&1.12&1.38&0.87&0.91&1.09\\
Br&1.31&2.41&0.84&0.80&1.34\\
Sr&1.26&1.60&0.88&0.86&1.21\\
Zr&1.20&1.40&0.88&0.91&1.12\\
Cd&1.11&1.40&0.80&0.86&0.96\\
Ba&1.26&1.69&0.84&0.82&1.16\\
La&1.22&1.54&0.85&0.86&1.13\\
Yb&1.20&1.63&0.83&0.83&1.12\\
Hg&1.09&1.33&0.74&0.86&0.85\\
Pb&1.10&1.22&0.77&0.95&0.89\\
Ra&1.22&1.64&0.91&0.81&1.21\\
\hline
\end{tabular}
\end{table}
Such an expression appears in atomic parity violating elecromagnetic amplitudes and, in addition to the weak
 matrix element, it contains the $E1$ amplitude and the energy
 denominator between the $s$ and the $p_{1/2}$ states. In molecules,
 use of this expression may be justified by the ionic bond model,
 where the electron that moves to the halogen polarizes the metal atom
 and produces a mixture of $s$ and $p$ orbitals.  
Table \ref{tab:IV} contains the obtained $K_W$ (corresponding to
$\braket{ns}{H_A}{np}$), along with $K_{E1}$ and $K_{E_n}$ factors,
which describe the effects of core polarization
(Eq. (\ref{eq:RPA})-(\ref{eq:meRPA})), and the correlations
(Eqs. \eqref{eq:Sigma}-\eqref{eq:K}) on the $E1$ amplitude and the
energy denominator, respectively. The final rescaling parameter $K_{tot}$ is
the product of all three factors
\begin{equation}
  K_{tot} = K_W \cdot K_{E1} \cdot K_{E_n}.
\label{eq:New_K}
\end{equation}

The final $K_{tot}$ factors can be found in Table \ref{tab:IV}, and are used
to scale the DHF $W_A$ parameters to include the effects of electron
correlation.  
(Note that that for PbF we have  calculated only correlations between the valence electrons and the Pb core;  the correlations between the valence electrons $6s$  and $6p$  are not included and  should be treated separately using a different technique).
The remaining correlation corrections, which are often called the weak
correlation potential $\delta \hat \Sigma_{\rm A}$ or structural
radiation, are suppressed by a small ratio $\epsilon_v/\epsilon_c \sim
0.1$, where $\epsilon_v$ is the valence electron energy and
$\epsilon_c$ is the core electron energy. The effect of 
$\delta \hat \Sigma_{\rm A}$ usually does not exceed a few
percent~\cite{DzuFlaSil87_2,DzuFlaSus89}.
 



\bibliography{anapoles}

\end{document}